\documentclass[prd]{revtex4}
\usepackage{amsmath}
\usepackage{epsfig}
 \usepackage{axodraw}

\begin{document}
\title{Determining the CP Violation Angle $\gamma$
in $B_s$ Decays without Hadronic Uncertainty}\thanks{This
work is partly supported by National Science Foundation of China.
}

\author{Bi-Hai Hong}
\affiliation{Department of Physics, Lishui University, Lishui,
Zhejiang 323000, China \\
 Department of Physics, Shanghai Jiaotong
University, Shanghai 200030, China}

\author{Cai-Dian L\"u}
\affiliation{  CCAST (World Laboratory), P.O. Box 8730, Beijing
100080, China \\
Institute of High Energy Physics, CAS, P.O.Box 918,  Beijing
100049, China\footnote{Mailing address.}}

\begin{abstract}
 We study the rare decays
$B_s^0\to D^\pm \pi^\mp$ and $\bar B_s^0\to D^\mp \pi^\pm$, which
can occur only via annihilation type $W$ exchange diagrams in the
standard model. The time-dependent decay rates of the four
channels can provide four CP parameters, which are experimentally
measurable. We show that the CKM angle $\phi_3=\gamma$ can be
determined from these parameters without any theoretical model
dependence. These channels can be measured in future LHCb
experiments to provide a clean way for $\gamma$ measurement.
\end{abstract}
%\newpage

 \maketitle

\section{Introduction}\label{s1}

The CP violation study  is one of the hot topics  in particle
physics. After the measurement of CKM angle $\phi_1=\beta$ in B
factories \cite{beta}, more attention has been drawn to the
extraction of the other two angles, especially $\phi_3 =\gamma$,
which is the most difficult one \cite{ligeti}.
 Besides the method based on approximation of SU(3) \cite{su3}, a lot of
other channels are discussed to measure this CKM angle, such as
$B\to DK$ decays \cite{glw}, $B\to K_S \pi^+\pi^-$ \cite{bkspp}
and
 $B\to D^{**} K$ decays \cite{bdkpp} etc. Most of the methods
 require a number of measurements, some require measurements of rare
 decays with small branching ratio. Therefore the measurement of
 angle $\gamma$ is still difficult for experiments.

 In this paper, we give another example to measure the CKM angle $\gamma$,
 which does not require any theoretical assumption, namely
 the rare decays
$B_s^0\to D^\pm \pi^\mp$ and $\bar B_s^0\to D^\mp \pi^\pm$.
Similar to the $B\to DK$ decays, there are both contributions from
$b\to c \bar u s$ and $b\to u \bar c s$ transitions, in these four
modes. The interference between the two kinds of decay amplitudes
will give out the information of CKM angle $\gamma$.   Unlike the
$B\to DK$ decays, there is no direct CP violation here, but mixing
induced CP violation, since neutral $B_s$ meson decays are
involved. The time dependent measurement of decay amplitudes can
provide the ratio of two decay amplitudes and the CKM angle
$\gamma$, without theoretical input.

Similar argument has also been proposed for $B_s (\bar B_s) \to
D_s^\pm K^\mp$ and $B^0 (\bar B^0) \to D^\pm \pi^\mp$ decays
\cite{alex} some years ago, which is intensively discussed later
in \cite{fleischer}. These
 decays   with emission diagram contributions
will have a larger branching ratio than the channels discussed
here. However the latter channels of $B^0(\bar B^0)$ decays
involve CKM matrix elements of $V_{ub}V_{cd}^*$ and
$V_{cb}V_{ud}^*$. The large difference between these two matrix
elements $|V_{ub}V_{cd}^*|\ll |V_{cb}V_{ud}^*|$ makes the two
decay amplitudes differ too much, thus experimentally too
difficult to measure. The $B_s (\bar B_s) \to D_s^\pm K^\mp$
decays should be the best channels to measure CKM angle $\gamma$
\cite{alex}. Our newly proposed channels $B \to D^\pm \pi^\mp$
will be an alternative choice.

\section{CP Asymmetry Variables of $B_s^0 (\bar B_s^0)\to D^\mp \pi^\pm$}\label{s2}

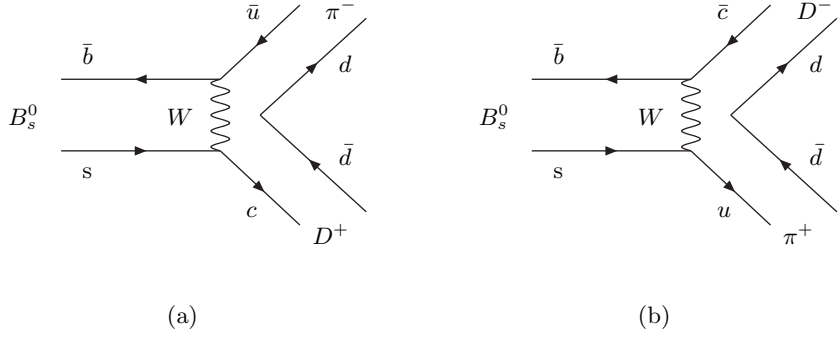
\begin{figure}[htbp]
   \begin{picture}(130,120)(0,-20)
        \ArrowLine(60,72)(0,72)
        \ArrowLine(0,45)(60,45)
        \ArrowLine(90,100)(60,72)
        \ArrowLine(60,45)(90,17)
        %\Gluon(30,20)(30,60){5}{4} \Vertex(30,20){1.5} \Vertex(30,60){1.5}
%------- weak vertex --------
        \Photon(60,45)(60,72){3.5}{4.5}
%------- weak vertex --------
        \ArrowLine(115,22)(75,58.5)
        \ArrowLine(75,58.5)(115,95)
       \put(70,95){$\bar u$}
      \put(-20,55){$B_s^0$}
      \put(100,95){$\pi^-$}
      \put(95,10){$D^+$}
      \put(105,40){$\bar d$}
      \put(105,75){$  d$}
          \put(70,20){$c$}
           \put(5,78){ {$\bar{b}$}}
      \put(5,35){ {s}}
       \put(40,55){$W$}
      \put(40,-20){(a)}
   \end{picture}
   \begin{picture}(110,120)(0,-20)
      \put(45,0){
        \ArrowLine(60,72)(0,72)
        \ArrowLine(0,45)(60,45)
        \ArrowLine(90,100)(60,72)
        \ArrowLine(60,45)(90,17)
        %\Gluon(30,20)(30,60){5}{4} \Vertex(30,20){1.5} \Vertex(30,60){1.5}
%------- weak vertex --------
        \Photon(60,45)(60,72){3.5}{4.5}
%------- weak vertex --------
        \ArrowLine(115,22)(75,58.5)
        \ArrowLine(75,58.5)(115,95)
       \put(70,95){$\bar c$}
      \put(-20,55){$B_s^0$}
      \put(100,95){$D^-$}
      \put(95,10){$\pi^+$}
      \put(105,40){$\bar d$}
      \put(105,75){$  d$}
          \put(70,20){$u$}
           \put(5,78){ {$\bar{b}$}}
      \put(5,35){ {s}}
       \put(40,55){$W$}
      \put(40,-20){(b)}}
   \end{picture}
 \caption{Perturbative Feynman diagrams contributing to decay
$B_s^0 \to D^+ \pi^-$ (a) and $B_s^0 \to D^- \pi^+$ (b).}
\label{fig1}
\end{figure}

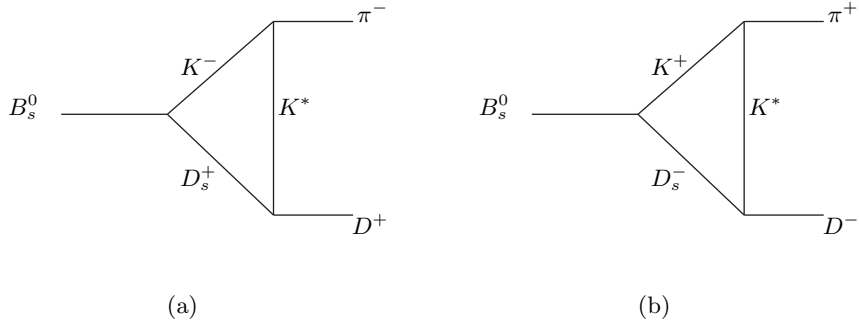
\begin{figure}[htbp]
   \begin{picture}(130,120)(0,-20)
        \Line(10,55)(50,55)
        \Line(50,55)(90,90)
        \Line(50,55)(90,17)
 %------- weak vertex --------
        \Line(90,90)(120,90)
        \Line(90,17)(120,17)
        \Line(90,17)(90,90)
       \put(55,70){$K^-$}
       \put(55,28){$D_s^+$}
      \put(-10,55){$B_s^0$}
      \put(92,55){$K^*$}
      \put(122,90){$\pi^-$}
      \put(120,10){$D^+$}
      \put(50,-20){(a)}
   \end{picture}
   \begin{picture}(130,120)(0,-20)
      \put(45,0){  \Line(10,55)(50,55)
        \Line(50,55)(90,90)
        \Line(50,55)(90,17)
 %------- weak vertex --------
        \Line(90,90)(120,90)
        \Line(90,17)(120,17)
        \Line(90,17)(90,90)
       \put(55,70){$K^+$}
       \put(55,28){$D_s^-$}
      \put(-10,55){$B_s^0$}
      \put(92,55){$K^*$}
      \put(122,90){$\pi^+$}
      \put(120,10){$D^-$}
      \put(50,-20){(b)}}
   \end{picture}
 \caption{Non-perturbative  diagrams contributing to decay
$B_s^0 \to D^+ \pi^-$ (a) and $B_s^0 \to D^- \pi^+$ (b).}
\label{fig2}
\end{figure}

 The non-leptonic $B_s^0$ decays $B_s^0\to D^+ \pi^-$ and $B_s^0\to D^-
 \pi^+$ are rare decays, which can occur only at tree level operators. No
 penguin operators can contribute to avoid the penguin pollution.
 The perturbative diagrams for these decays are shown in Figure
 \ref{fig1}.
 For decay $B_s^0\to D^+ \pi^-$ (Fig.\ref{fig1}(a)), the decay amplitude is proportional to
$ V_{ub}^*V_{cs}$.
  And for decay $B_s^0\to D^- \pi^+$ (Fig.\ref{fig1}(b)),
the decay amplitude is proportional to $V_{cb}^* V_{us}$. They are
pure annihilation type decays with W exchange diagrams. Despite
the perturbative picture, one may argue that they can get
contribution from non-perturbative diagrams, such as soft-final
state interaction diagrams shown in Fig.\ref{fig2}: $B_s^0 \to
D_s^\pm K^\mp \to D^\pm \pi^\mp$. Fortunately, these diagrams have
the same CKM matrix elements with the perturbative one in
Fig.\ref{fig1} to make these channels clean for CP violation
measurement. Therefore, the decay amplitudes for these decays can
be parameterized as
\begin{equation}\begin{array}{ll}
g=\langle D^+\pi^-|H|B_s^0\rangle ~= V_{ub}^*V_{cs} A_1,&
h=\langle D^+\pi^-|H|\bar B_s^0\rangle~= V_{cb}V_{us}^* A_2,\\
\bar g=\langle D^-\pi^+|H|\bar B_s^0\rangle~= V_{ub}V_{cs}^* A_1,
& \bar h=\langle D^-\pi^+|H|B_s^0\rangle~=
V_{cb}^*V_{us}A_2,\label{amp}
\end{array}\end{equation}
 which determine the decay matrix elements of
$B_s^0\to D^+\pi^-$ and $D^-\pi^+$, and of $\bar B_s^0\to
D^-\pi^+$ and $D^+\pi^-$.  There is only one kind of contribution
for each of the decay modes, thus there is no direct CP violation
for them. However there is still a CP violation induced by mixing,
although they are decays with charged final states (non CP
eigenstates).

The neutral $B_s -\bar B_s$ mixing is usually described as
\begin{gather}
B_H=p|B_s^0>+q|\bar B_s^0>,\\
B_L=p|B_s^0>-q|\bar B_s^0>,
\end{gather}
with $|p|^2+|q|^2=1$. $|p|$ and $|q|$ come from the information of
$B_s^0$ and $\bar B_s^0$ transition from each other. Thus,
\begin{gather}
\frac{q}{p}=\frac{V_{tb}^* V_{ts}}{V_{tb} V_{ts}^*} ,
\end{gather}
for $B_s$-$\bar B_s$ mixing. In Wolfenstein parameterization
\cite{wolfenstein}, $arg(q/p) =2\lambda^2\eta< 2^\circ$, which is
negligible.  Since we are only interested in the CKM angle
measurement, the normalized time-dependent decay rates for $B_s
\to D^{\pm}\pi^{\mp}$ are given by \cite{alex2}:
\begin{eqnarray}
  \Gamma^{ D^\pm\pi^\mp} (t)
& =& (1\pm A_{CP}) \frac{e^{-t/\tau_{B_s}}}{8\tau_{B_s}} \left\{
1+   (S_{D\pi} \pm \Delta S_{D\pi}) \sin \Delta m t \right. \nonumber\\
&&\left.  + (C_{D\pi} \pm \Delta C_{D\pi})\cos \Delta m t \right
\},
  \label{rate}
\end{eqnarray}
and $ \bar B_s \to D^{\pm}\pi^{\mp}$ by
\begin{eqnarray}
  \bar \Gamma^{ D^\pm\pi^\mp} (t)
& =& (1\pm A_{CP}) \frac{e^{-t/\tau_{B_s}}}{8\tau_{B_s}} \left\{
1-   (S_{D\pi} \pm \Delta S_{D\pi}) \sin \Delta m t \right. \nonumber\\
&&\left.- (C_{D\pi} \pm \Delta C_{D\pi})\cos \Delta m t \right \},
  \label{rate2}
\end{eqnarray}
where $\Delta m$ is the mass difference of the two mass
eigenstates $B_H$ and $B_L$, and
\begin{equation}\begin{array}{ll}
C_{D\pi} = \frac{1}{2} (a_{\epsilon' }+a_{\bar \epsilon '}) ,&
\Delta C_{D\pi} =\frac{1}{2} (a_{\epsilon'}-a_{\bar\epsilon'}),\\
 S_{D\pi} =\frac{1}{2} (a_{\epsilon +\epsilon '}+a_{\epsilon +\bar \epsilon '}) ,&
 \Delta S_{D\pi}
=\frac{1}{2} (a_{\epsilon +\epsilon '}-a_{\epsilon +\bar \epsilon
'}).
\end{array}\end{equation}
They can be expressed by another set of parameters as
\begin{equation}
  \label{aepsilon}
  \begin{array}{ll}
a_{\epsilon '} = \displaystyle \frac{ |g|^2 -|h|^2}{ |g|^2
+|h|^2}, & a_{\epsilon +\epsilon '} = \displaystyle \frac{-2Im
\left(  {h}/{g}\right)}
{1+|h/g|^2},\\
a_{\bar \epsilon '} = \displaystyle \frac{ |\bar h|^2 -|\bar
g|^2}{ |\bar h|^2 +|\bar g|^2},
 &
a_{\epsilon +\bar \epsilon '} = \displaystyle \frac{-2Im \left(
 {\bar g} /{\bar h}\right)} {1+|\bar g/\bar h |^2}.
   \end{array}
\end{equation}
Utilizing eq.(\ref{amp}), we can get
\begin{equation}\begin{array}{ll}
C_{D\pi} =A_{CP}=0 ,&
\Delta C_{D\pi} =\displaystyle\frac{ 1 -R^2}{ 1 +R^2},\\
 S_{D\pi} =\displaystyle \frac{2 R \sin \gamma \cos \delta} {1+R^2} ,&
 \Delta S_{D\pi}
=\displaystyle \frac{-2 R \sin \delta \cos \gamma} {1+R^2} ,
\end{array}\label{cpf}
\end{equation}
where $R=|h/g|= |V_{cb}V_{us}^* A_2|/|V_{ub}^*V_{cs} A_1| $, is
the relative size of the two kinds of decay amplitudes. And
$\delta=arg(A_2/ A_1)$ is the relative strong phase between them.
 From eq.(\ref{cpf}), one can easily see that the ratio $R$ can be
determined from $\Delta C_{D\pi}$, and strong phase $\delta$ and
the CKM angle $\phi_3=\gamma$ can be solved from $S_{D\pi}$ and
$\Delta S_{D\pi}$, without uncertainty, if $\Delta
C_{D\pi}$,$S_{D\pi}$ and $\Delta S_{D\pi}$ have been gotten from
experiments. From eq.(\ref{rate},\ref{rate2}), one can also see
that, $\Delta C_{D\pi}$,  $S_{D\pi}$ and $\Delta S_{D\pi}$ are
measurable by experiments through the time-dependent decay rate.
In the standard model (SM), the strong phase of $A_1$ and $A_2$
should be the same, since CP
 is conserved in strong interaction. Therefore, $\delta=0$, and $ \Delta
 S_{D\pi}=0$. Hence only $\Delta C_{D\pi} $ and
 $  S_{D\pi}$ are used to determine $R$ and $\sin \gamma$. In a word, the CKM angle
$\phi_3=\gamma$ can be determined cleanly without any theoretical
model dependence, provided the experimental measurements of
time-dependent decay rates.

The parameter $C_{D\pi} =A_{CP}= \Delta
 S_{D\pi}=0$ is a consequence of the fact
that there is only one kind of contribution for each of the
decays. If there is any new physics contribution, which usually
provides a different weak phase, these two parameters will not be
zero any longer. The non-zero measurement of these parameters
experimentally will be a signal of new physics.

Since these decays are rare decays, one may worry about the decay
branching ratios are too small to be measured. A perturbative QCD
approach (PQCD) based on $k_T$ factorization shows that they are
at least  at the order of $10^{-6}$ \cite{liying}. This is
consistent with naive argument that the annihilation topology is
power suppressed as $1/m_b$, which is order of 10\%. Translating
to branching ratios, the $B_s\to D\pi$ branching ratio ($10^{-6}$)
should be at 1\% level of the emission type decay $B_s\to D_s K$
($10^{-4}$).
 The CP
  parameters $S_{D\pi}$ and $\Delta S_{D\pi}$ are  also
sensitive to the relative amplitude $R$ through eq.(\ref{cpf}). If
$R$ is too small or too big, $S_{D\pi}$ and $\Delta S_{D\pi}$ will
be too small to be measured. The same study in PQCD shows $R\simeq
1.8$ \cite{liying}, to make the extraction of CKM angle
$\phi_3=\gamma$ realistic. In fact, the ratio $R$ can not deviate
from 1 too much, since the CKM parameter $|V_{cb}V_{us}^*|$ and
$|V_{ub}^*V_{cs} | $ for these two kinds of decays are at the same
level  ${\cal O} (\lambda^3)$ in Wolfenstein parameterization
\cite{wolfenstein}. More precisely, the value of
${|V_{ub}^*V_{cs}|}$ is about half of ${|V_{cb}^*V_{us}|}$.

The same argument shown above is applicable to the $B_s (\bar B_s)
\to D_s^\pm K^\mp$ decays. Since there are emission diagram
contributions for these decays, their decay branching ratios are
much higher at order of $10^{-4}$ \cite{liu}. It is easier for
experiments to measure. The branching ratio of the proposed
channel $ B_s(\bar B_s )\to D^{\pm}\pi^{\mp}$ is two order
magnitude smaller, but it will provide a test for SM to measure
the same quantity using different channels. The situation is
similar in the $\beta$ measurement, where people try to measure CP
asymmetry of $B\to K_s \phi$, after $B\to J/\psi K_s$. Recently,
many new physics discussions have been made on this issue due to
the different results for the two channels \cite{phik}.

Since the current B factories do not produce $B_s$ mesons, there
are no data for these
 decays in experimental side up to now. But the designed LHCb
experiment will produce   $10^{12}$ $b\bar b$ pairs per year where
10\% of them will be  $B_s (\bar B_s)$ \cite{chang}. The $B_s-
\bar B_s$ mixing parameter $\Delta m_{B_s}$, which is predicted to
be 25 ps$^{-1}$ in SM, can be measured in LHCb in one month. With
$10^{11}$ $B_s$ mesons produced, the LHCb experiment can measure
decays with branching ratio as small as $10^{-7}$ even if the
detection efficiency is only several percent. Therefore the $
B_s(\bar B_s )\to D^{\pm}\pi^{\mp}$ decay is measurable in the
near future, although the time-dependent CP asymmetry measurement
could be challenging.

The $B^0 (\bar B^0) \to D^\pm \pi^\mp$ decays, can be easily
measured  in the current B factories, however, the same argument
does not apply to them, since its ratio $R\simeq |V_{ub}V_{cd}^*|/
|V_{cb}^*V_{ud} |\simeq 0.02$  is too small, making the
measurement of $S_{D\pi}$ and $\Delta S_{D\pi}$ nonrealistic.

\section{Summary} \label{s4}

In this paper, we show that the four time-dependent decay rates of
$B_s^0\to D^\pm \pi^\mp$ and $\bar B_s^0\to D^\mp \pi^\pm$, can
provide four CP parameters which are experimentally measurable.
These parameters are functions of CKM angle $\phi_3=\gamma$. The
measurement of these parameters at the future LHCb experiment can
provide a method to extract CKM angle $\gamma$ without any
hadronic uncertainty.  These channels occur purely via
annihilation type $W$ exchange  diagrams, and have a branching
ratio of $10^{-6}$ which is measurable in future LHCb experiment.

\section*{Acknowledgement}

 We thank Y.N. Gao for discussions on the
LHCb experiments.

\end{document}